\documentclass[amsmath,amssymb,12pt,superscriptaddress,nofootinbib]{revtex4}

\usepackage{graphicx}
\usepackage{dcolumn}
\usepackage{bm}
\usepackage{color}
\usepackage{enumitem}
  \usepackage{tabularx}
   \newcolumntype{C}{>{\centering\arraybackslash}X}
   \newcolumntype{L}{>{\raggedright\arraybackslash}X}
   \newcolumntype{R}{>{\raggedleft\arraybackslash}X}
\setlistdepth{10}

\usepackage[normalem]{ulem}

\newcommand{\ii}{\mathrm{i}}
\newcommand{\dd}{\mathrm{d}}

\newcommand{\del}{\partial}
\newcommand{\ee}{{\rm e}}

\definecolor{DarkBlue}{rgb}{0,0,0.7} 

\definecolor{DarkRed}{rgb}{0.65,0,0}

\begin{document}
\baselineskip5.5mm

{\baselineskip0pt
\small
\leftline{\baselineskip16pt\sl\vbox to0pt{
                             \vss}}
\rightline{\baselineskip16pt\rm\vbox to20pt{
\vspace{-1.5cm}
            \hbox{YITP-21-23}
\vss}}
}

\author{Chul-Moon~Yoo}\email{yoo@gravity.phys.nagoya-u.ac.jp}

\affiliation{
\fontsize{12pt}{1pt}\selectfont
Division of Particle and Astrophysical Science,
Graduate School of Science, \\Nagoya University, 
Nagoya 464-8602, Japan
\vspace{1.5mm}
}

\author{Atsushi~Naruko}

\affiliation{
\fontsize{12pt}{1pt}\selectfont
Center for Gravitational Physics, Yukawa Institute for Theoretical Physics, Kyoto University, Kyoto 606-8502, Japan
\vspace{1.5mm}
}

\author{Yusuke~Sakurai}

\affiliation{
\fontsize{12pt}{1pt}\selectfont
Division of Particle and Astrophysical Science,
Graduate School of Science, \\Nagoya University, 
Nagoya 464-8602, Japan
\vspace{1.5mm}
}

\author{Keitaro~Takahashi}

\affiliation{
\fontsize{12pt}{1pt}\selectfont
Kumamoto University, Graduate School of Science and Technology, Kumamoto, 860-8555, Japan
\vspace{1.5mm}
}

\affiliation{
\fontsize{12pt}{1pt}\selectfont
International Research Organization for Advanced Science and Technology, Kumamoto University, Kumamoto 860-8555, Japan
\vspace{1.5mm}
}

\affiliation{
\fontsize{12pt}{1pt}\selectfont
National Astronomical Observatory of Japan, 2-21-1 Osawa, Mitaka, Tokyo 181-8588, Japan
\vspace{1.5mm}
}

\author{Yohsuke~Takamori}

\affiliation{
\fontsize{12pt}{1pt}\selectfont
National Institute of Technology (KOSEN), Wakayama College, Gobo, Wakayama 644-0023, Japan
\vspace{1.5mm}
}

\author{Daisuke~Yamauchi}

\affiliation{
\fontsize{12pt}{1pt}\selectfont
Faculty of Engineering, Kanagawa University, Kanagawa-ku, Yokohama-shi, Kanagawa, 221-8686, Japan
\vspace{1.5mm}
}

\vskip-1.cm
\title{Axion Cloud Decay due to the Axion-photon Conversion 
\\with Background Magnetic Fields}


\begin{abstract}
\baselineskip5.5mm 
We consider an axion cloud around a black hole with 
background magnetic fields. 
We calculate the decay rate of the axion cloud 
due to the axion-photon conversion associated with 
the axion-photon coupling. 
For simplicity, we consider the situation where the 
axion configuration is dominated by a solution for 
the eigenvalue equation equivalent to that for the Hydrogen atom, and  
the coupling term can be evaluated by a successive perturbation method. 
For the monopole background, 
we find the decay rate of the axion cloud is given by 
$\sim q^2\kappa^2(GM)^5\mu^8$, where 
$\mu$, $M$, $G$, $\kappa$ and $q$ are 
the axion mass, black hole mass, gravitational constant, 
coupling constant of the axion-photon coupling and monopole charge, 
respectively. 
For the uniform background magnetic field, we obtain 
the decay rate of the axion cloud 
$\sim B_0^2\kappa^2 (GM)^7\mu^6$, where $B_0$ is 
the magnetic field strength. 
Applying our formula to the central black hole in our galaxy, 
we find that the value of the decay rate for the case of the uniform magnetic field 
is comparable to the growth rate of the superradiant instability 
with $\kappa\sim 10^{-12}{\rm GeV^{-1}}$, 
$B_0\sim 10^3{\rm G}$ and $\mu\sim 10^{-18}{\rm eV}$. 
The ratio is $10^5$ times larger for the monopole magnetic field 
with the same values of the parameters. 
\end{abstract}


\maketitle
\thispagestyle{empty}
\pagebreak

\section{Introduction}

Recently, axion-like particles(ALPs) have been attracting much interest 
in cosmology, astrophysics and particle physics. 
The axion was originally proposed as a solution of the strong CP problem 
in quantum chromodynamics~(QCD)~\cite{Peccei:1977hh,Peccei:1977ur,Weinberg:1977ma,Wilczek:1977pj}, 
and the possibility of ALPs in string theory was pointed out in late years~\cite{Svrcek:2006yi,Arvanitaki:2009fg}. 
ALPs including the QCD axion are attractive candidates for dark matter~\cite{Preskill:1982cy,Dine:1982ah,Abbott:1982af}. 
Various experimental and observational searches for ALPs are ongoing, 
and the constraints are actively updated~(see, e.g. \cite{Zyla:2020zbs} for a summary of constraints).

One interesting phenomenon associated with the existence of an ALP is 
formation of the axion cloud surrounding a black hole~\cite{Arvanitaki:2010sy}. 
For a massive bosonic field around a rotating black hole, 
superradiant instability takes place\cite{Zouros:1979iw,Detweiler:1980uk,Brito:2015oca}, 
and the specific modes of the massive field exponentially grow with time. 
The growing time scale for the massive scalar field is analytically estimated in Refs.~\cite{Zouros:1979iw,Detweiler:1980uk} 
under certain approximations and numerically calculated in Ref.~\cite{Dolan:2007mj}. 
Possible observable phenomena and those relevance have been discussed in a number of papers~\cite{Cardoso:2011xi,Yoshino:2012kn,Yoshino:2013ofa,Brito:2014wla,Yoshino:2015nsa,Arvanitaki:2014wva,Brito:2017zvb,Baumann:2018vus,Zhang:2018kib,Zhang:2019eid,Baumann:2019ztm,Ding:2020bnl,Omiya:2020vji}. 
If the Compton wavelength of an ALP is similar to the gravitational radius of a rotating black hole, 
the angular momentum of the black hole may be efficiently extracted and exhausted by the superradiant instability
within the cosmic history. 
Therefore the existence of a rotating black hole may give a constraint on the existence of an ALP 
in the corresponding mass range~\cite{Stott:2020gjj}. 
However, once we consider more general situations, 
there may be an efficient decaying process which spoils the persistent growth of the cloud. 
For instance, the existence of a binary compact object 
may dissolve the cloud due to the level transitions~\cite{Takahashi}. 
In this paper, 
we focus on the dissipative effect through 
the coupling with the electro-magnetic field.

An ALP may have a coupling with the electro-magnetic field as the Chern-Simons coupling. 
Many axion search experiments rely on this coupling and gave constraints on 
the magnitude of the coupling 
constant~\cite{Anastassopoulos:2017ftl,Boutan:2018uoc,Ouellet:2018beu,Calore:2020tjw,Salemi:2021gck}. 
The birefringence due to this coupling term under the existence of the ALP is also actively discussed~\cite{Carroll:1989vb,Carroll:1991zs,Harari:1992ea,Ivanov:2018byi,Fujita:2018zaj,Liu:2019brz,Fedderke:2019ajk,Caputo:2019tms,Chen:2019fsq,Yuan:2020xui,Basu:2020gsy}. 
Another characteristic phenomenon is the axion-photon conversion with a background magnetic field~\cite{Maiani:1986md,Raffelt:1987im}, 
that is, the axion field can be converted to electro-magnetic waves through the coupling with the background magnetic field. 
There is a number of related papers to the axion-photon conversion~\cite{Sikivie:1983ip,Sikivie:1985yu,Ahonen:1995ky,Pshirkov:2007st,Espriu:2011vj,Espriu:2014lma,Huang:2018lxq,Hook:2018iia,Hertzberg:2018zte,Arza:2018dcy,Masaki:2019ggg,Carenza:2019vzg,Arza:2020eik,Leroy:2019ghm,Buckley:2020fmh}. 
In this paper, we discuss the effect of the axion-photon conversion on the growth/decay rate of the axion cloud.

Let us consider a rotating black hole surrounded by a magnetic field and an ALP cloud. 
While the ALP has unstable bound states without the electro-magnetic field, 
if the ALP has the coupling with the electro-magnetic field, 
the axion-photon conversion takes place and the photon can dissipate to infinity~(see also Ref.~\cite{Boskovic:2018lkj} 
for effects of a magnetic field on 
an axion cloud). 
The decay due to the axion-photon conversion and its phenomenological implications are also considered in Ref.~\cite{Blas:2020nbs}. 
Therefore we expect the competing effects of the superradiant growth and the decay due to the axion-photon conversion.%
\footnote{
\baselineskip5mm
The effect of a magnetic field on the axion field through the modification of the background geometry
 is discussed in Ref.~\cite{Brito:2014nja}, 
and the charged scalar field in a magnetized black hole is discussed in Ref.~\cite{Turimov:2019afv}. 
However, as far as we know, the effect of the axion-photon conversion to the stability of 
the axion cloud has not been reported yet. 
}
As a first step, we consider the 
case in which the typical size of the cloud is much larger than the gravitational radius of the black hole. 
Then the spin of the black hole and the absorption through the horizon 
can be neglected, that is, the black hole can be treated as a gravitational source of the point mass. 
We calculate the decay rate due to the axion-photon conversion 
for the dominant mode of the superradiant instability assuming the axion-photon coupling term is perturbatively small. 
We note that, although the spin of the black hole is essential for the superradiant instability, 
as will be shown in the text, 
the effect of the axion-photon conversion can be estimated independently of the spin of the black hole 
once the dominant mode of the axion cloud is specified.  
The magnetic field configuration is assumed to be the monopole in Sec.~\ref{sec:mono} and 
uniform~\cite{Wald:1974np} in Sec.~\ref{sec:wald}. 
Finally we compare the decay rate to the growth rate of the superradiant instability in Sec.~\ref{sec:comp}.

Throughout this paper, we use the natural units in which both 
the speed of light and the reduced Planck constant are unity, $c=\hbar=1$, 
and the gravitational constant is denoted by $G$.

\section{Equations of motion, background and perturbations}
\label{sec:mono}

Let us consider the axion-electro-magnetic system given by the following action:
\begin{equation}
S=\int\sqrt{-g}\dd^4x\left(-\frac{1}{4}F_{\mu\nu}F^{\mu\nu}-\frac{1}{4}\kappa \phi F_{\mu\nu}\tilde F^{\mu\nu}-\frac{1}{2}\nabla_\mu\phi \nabla^\mu \phi
-\frac{1}{2}\mu^2\phi^2\right), 
\end{equation}
where
\begin{equation}
\tilde F^{\mu\nu}=\frac{1}{2}\varepsilon^{\mu\nu\lambda\rho}F_{\lambda\rho}
\end{equation}
with $\varepsilon$ being the Levi-Civita tensor, and we neglected the non-linear self-interaction of the axion field. 
From the variation with respect to the axion field, we obtain the following equation of motion for the axion:
\begin{equation}
\left(\nabla_\mu\nabla^\mu-\mu^2\right)\phi=\frac{1}{4}\kappa F_{\mu\nu}\tilde F^{\mu\nu}. 
\label{eq:axioneom}
\end{equation}
The equations of motion for the gauge field are given by 
\begin{equation}
\nabla_\mu F^{\mu\nu}=-\kappa\tilde F^{\mu\nu}\nabla_\mu \phi, 
\label{eq:Maxwell}
\end{equation}
where we have used the following identity:  
\begin{equation}
\nabla_\mu\tilde F^{\mu\nu}=\frac{1}{2}\varepsilon^{\mu\nu\rho\lambda} \nabla_{\mu} F_{\rho\lambda}=0. 
\end{equation}
For the background geometry, we consider the Schwarzschild metric with the mass $M$ given by 
\begin{equation}
\dd s^2=-f(r)\dd t^2+\frac{\dd r^2}{f(r)}+r^2(\dd \theta^2+\sin^2\theta \dd \varphi^2), 
\end{equation}
where $f(r)=1-2GM/r$. 

In the following sections, we will consider perturbations on the two types of background magnetic field configurations 
satisfying Eq.~\eqref{eq:Maxwell} with $\phi=0$. 
That is, the axion $\phi$ and the gauge fields $A^{\rm tot}_\mu$ will given in the following form: 
\begin{eqnarray}
\phi&=&\delta \phi,\\
A^{\rm tot}_\mu&=&A^{\rm bg}_\mu +\delta A_\mu, 
\end{eqnarray}
where $A^{\rm bg}$ is the background gauge field satisfying Eq.~\eqref{eq:Maxwell} with $\phi=0$, 
and $\delta \phi$ and $\delta A_\mu$ are perturbations. 
We will consider the equations of motion for $\delta \phi$ and $\delta A_\mu$ 
at the linear order. 

In the form of the vector spherical harmonics~\cite{Zerilli:1974ai} and the Fourier mode expansion with the frequency $\omega$, 
we can expand the axion field and each component of the gauge field as 
\begin{eqnarray}
\delta\phi&=&\sum_{lm} \Phi_{lm} Y_{lm}\ee^{\ii \omega t}, \\
\delta A_t&=&-\ii\sum_{lm} A^a_{lm} Y_{lm}\ee^{\ii \omega t}, 
\label{eq:AtY}
\\
\delta A_r&=&\sum_{lm} A^b_{lm} Y_{lm}\ee^{\ii \omega t}, 
\label{eq:ArY}\\
\delta A_\theta&=&\sum_{lm} \frac{1}{\sqrt{l(l+1)}}\left(A^c_{lm}\del_\theta Y_{lm}+A^d_{lm}\frac{1}{\sin\theta}\del_\varphi Y_{lm}\right)\ee^{\ii \omega t},
\label{eq:AthY}\\
\delta A_\varphi&=&\sum_{lm} \frac{1}{\sqrt{l(l+1)}}\left(A^c_{lm}\del_\varphi Y_{lm}-A^d_{lm}\sin\theta \del_\theta Y_{lm}\right)\ee^{\ii \omega t}, 
\label{eq:AphY}
\end{eqnarray}
where, $Y_{lm}$ is the spherical harmonic function of degree $l$ and order $m$, 
and $\Phi_{lm}$, $A^a_{lm}$, $A^b_{lm}$, $A^c_{lm}$ and $A^d_{lm}$ are functions of $r$. 
The signs of $A^a_{lm}$, $A^b_{lm}$ and $A^c_{lm}$ change as 
$(-1)^l$ for the parity transformation $(\theta, \varphi)\rightarrow(-\theta, \varphi+\pi)$ 
and the sign of $A^d_{lm}$ changes as $(-1)^{l+1}$, and they are called even and odd parity modes, respectively.

\section{Analysis with the monopole magnetic field}
\label{sec:mono}

\subsection{perturbation equations}
First, as the simplest case, 
let us consider the following perturbation around the monopole magnetic field:
\begin{eqnarray}
A^{\rm tot}_\mu&=&q(1-\cos\theta)(\dd\varphi)_\mu +\delta A_\mu. 
\end{eqnarray}
Since the monopole field does not violate the spherical symmetry, 
the odd parity modes and the even parity modes are decoupled with each other, and 
only the even parity modes can be coupled with the axion field equation. 
Since each mode labelled by $lm$ decouples from the other modes, 
we just focus on the single mode specified by $l$ and $m$, 
and omit the labels $l$ and $m$ for notational simplicity in this section. 

We fix the gauge as $A^c=0$~\cite{Zerilli:1974ai}. 
The equations of motion for the gauge field are given by 
\begin{eqnarray}
l(l+1)A^a-f\del_r(r^2\del_r A^a)-\omega f\del_r(r^2 A^b)&=&-\ii fq\kappa\del_r \Phi, 
\label{eq:A1}\\
-fl(l+1)A^b+\omega r^2(\del_r A^a+\omega A^b)&=&\ii q\kappa\omega\Phi, \label{eq:A2}\\
f\del_r(f A^b)-\omega A^a&=&0. \label{eq:dtAt}
\end{eqnarray}
Substituting Eq.~\eqref{eq:dtAt} into Eq.~\eqref{eq:A2}, we find 
\begin{equation}
-fl(l+1)A^b+r^2\del_r\left[f\del_r(fA^b)\right]+r^2\omega^2A^b=\ii q\kappa\omega\Phi. 
\label{eq:fullA}
\end{equation}
Eq.~\eqref{eq:A1} is redundant. 
The axion field equation is given by 
\begin{equation}
\frac{f}{r^2}\del_r\left(r^2f\del_r \Phi\right)+\omega^2\Phi
-f\mu^2\Phi-f\frac{l(l+1)}{r^2}\Phi=\frac{1}{4}\kappa f F_{\mu\nu}\tilde F^{\mu\nu}, 
\label{eq:fullphieom}
\end{equation}
where $F_{\mu\nu}\tilde F^{\mu\nu}$ is given by 
\begin{equation}
F_{\mu\nu}\tilde F^{\mu\nu}=-\frac{4\ii q}{r^2}\left(\del_rA^a+\omega A^b\right)
=-\ii\frac{4q}{\omega r^2}\left[\del_r(f\del_r(fA^b))+\omega^2A^b\right]. 
\end{equation}
Therefore we obtain the two coupled equations \eqref{eq:fullA} and \eqref{eq:fullphieom} as master equations.

\subsection{Approximate Analysis in analogy with the Hydrogen atom}
In order to analytically evaluate the equations, we impose the following conditions: 
\begin{eqnarray}
&1/(GM)\gg\omega\sim\mu\gg1/r,
\label{eq:approx}
\\
&q\kappa\ll r, 
\label{eq:smallcharge}
\end{eqnarray}
where $r$ denotes the typical length scale of the system. 
The last condition indicates that the coupling terms are relatively small 
and can be evaluated by using a successive approximation. 
Under the condition \eqref{eq:approx}, 
the master equations \eqref{eq:fullA} and \eqref{eq:fullphieom}  can be reduced to
\begin{eqnarray}
\frac{\dd^2}{\dd r^2}A^b+\omega^2A^b-\frac{l(l+1)}{r^2}A^b&=&\ii\frac{\omega q\kappa}{r^2} \Phi, 
\label{eq:A}
\\
\frac{1}{r^2}\frac{\dd}{\dd r}\left(r^2\frac{\dd}{\dd r}\Phi\right)+\left(\omega^2-\mu^2+\frac{2GM}{r}\mu^2-\frac{l(l+1)}{r^2}\right)\Phi
&=&-\ii\frac{q\kappa}{\omega r^2}\left[\frac{\dd^2}{\dd r^2}A^b+\omega^2A^b\right].   
\label{eq:phi}
\end{eqnarray}
Combining these two equations, we can rewrite Eq.~\eqref{eq:phi} as   
\begin{equation}
\frac{1}{r^2}\frac{\dd}{\dd r}\left(r^2\frac{\dd}{\dd r}\Phi\right)+\left(\omega^2-\mu^2+\frac{2GM\mu^2}{r}-\frac{l(l+1)}{r^2}\right)\Phi
=\frac{q^2\kappa^2}{r^4}\Phi-\ii\frac{q\kappa l(l+1)}{\omega r^4} A^b. 
\label{eq:phi2}
\end{equation}

Hereafter we successively evaluate Eqs.~\eqref{eq:A} and \eqref{eq:phi2} 
setting the lowest order equation as 
\begin{equation}
\frac{\dd^2\Phi_0}{\dd r^2}+\frac{2}{r}\frac{\dd\Phi_0}{\dd r}+\left(\omega_0^2-\mu^2\right)\Phi_0
+\frac{2GM\mu^2}{r}\Phi_0-\frac{l(l+1)}{r^2}\Phi_0=0. 
\label{eq:Hydro}
\end{equation}
This equation is equivalent to the equation for the Hydrogen atom. 
The radius $a_0$ 
corresponding to the Bohr radius is given by 
\begin{equation}
a_0=1/(GM\mu^2). 
\end{equation}
This radius gives the typical radius of the system. 
Let us focus on the $l=m=1$ and $n=2$ mode since it has the largest growth rate in 
the context of the superradiant instability, where $n$ is the principal quantum number. 
The solution for $l=1$ and $n=2$ mode is given by 
\begin{equation}
\Phi_0=\omega_0\left(\frac{1}{2a_0\omega_0}\right)^{3/2}\frac{r}{\sqrt{3}a_0}\exp \left(-\frac{r}{2a_0}\right), 
\label{eq:phi21}
\end{equation}
and the frequency is given by 
\begin{equation}
\omega_0^2-\mu^2=-\frac{1}{4a_0^2}, 
\end{equation}
where we have normalized $\Phi_0$ as $\omega_0\int \dd r r^2 \Phi_0^2=1$ keeping the dimension of $\Phi_0$ as the mass dimension one.

As for Eq.~\eqref{eq:A},  
using the variable $\mathcal A:=A^b/(\omega_0 r)$, we obtain 
\begin{equation}
\frac{\dd^2 \mathcal A}{\dd r^2}+\frac{2}{r}\frac{\dd \mathcal A}{\dd r}+\omega_0^2\mathcal A-\frac{l(l+1)}{r^2}\mathcal A=\ii \frac{q \kappa}{r^3} \Phi_0, 
\end{equation}
where we neglected the higher order contributions, and replaced $\Phi$ and $\omega$ by $\Phi_0$ and $\omega_0$, respectively. 
By using the dimensionless variable $x:=\omega r$, this equation can be rewritten as 
\begin{equation}
\frac{\dd}{\dd x}\left(x^2\frac{\dd}{\dd x}\mathcal A\right)+x^2 \mathcal A-l(l+1)\mathcal A=\ii q\kappa \omega_0 x^{-1}\Phi_0. 
\end{equation}
The homogeneous solution can be described by the spherical Bessel functions $j_l(x)$ and $n_l(x)$. 
The solution which is regular at the origin is given by $j_l$. 
We note that the absorption through the black hole horizon has been neglected 
under the assumption \eqref{eq:approx}. 
For the boundary condition at infinity, since there is no outer potential barrier, 
we should take the outgoing boundary condition. 
Thus the behavior at infinity should be given by $\sim \ee^{-ix}$. 
This boundary condition can be satisfied by the second spherical Hankel function defined by $h_l(x):=j_l-\ii n_l$. 
Then the appropriate Green's function can be written as 
\begin{equation}
\mathcal G (x,\xi)=\Theta(\xi-x)j_l(x)h_l(\xi)/W+\Theta(x-\xi)j_l(\xi)h_l(x)/W, 
\label{eq:green}
\end{equation}
where $\Theta$ is the Heviside's step function and $W=x^2[\del_x j_l(x) h_l(x)-j_l(x)\del_x h_l(x))]$. 
For a given $\Phi_0$, the solution of $\mathcal A$ can be given by 
\begin{equation}
\mathcal A(x)=-\ii q\kappa \omega_0 \int^\infty_0 \dd\xi \mathcal G(x,\xi)\xi^{-1}\Phi_0(\xi). 
\end{equation}
For $l=1$, we have 
\begin{equation}
j_1(x)=\frac{\sin x}{x^2}-\frac{\cos x}{x},~h_1(x)=\frac{\ii-x}{x^2}\ee^{-\ii x}. 
\end{equation}
Then $W=\ii$ and the real part of $\mathcal A$ is given by 
\begin{equation}
{\rm Re}\mathcal A(x)=-q\kappa \omega_0 j_1(x)\int^\infty_0 \dd\xi j_1(\xi)\xi^{-1}\Phi_0(\xi). 
\label{eq:reA}
\end{equation}

The 1-st order equation for the axion field is given by 
\begin{equation}
\frac{\dd^2 \Phi_1}{\dd r^2}+\frac{2}{r}\frac{\dd \Phi_1}{\dd r}+\left(\omega_0^2-\mu^2\right)\Phi_1
+2\omega_0\omega_1\Phi_0
+\frac{2}{a_0r}\Phi_1-\frac{l(l+1)}{r^2}\Phi_1=\frac{q^2\kappa^2}{r^4}\Phi_0-\ii\frac{q\kappa l(l+1)}{r^3}\mathcal A. 
\label{eq:1stmono}
\end{equation}
Multiplying $r^2\Phi_0$ and integrating by parts, we obtain 
\begin{equation}
2\omega_0\omega_1\int^\infty_0\dd r r^2\Phi_0^2=q^2\kappa^2\int^\infty_0\dd r\frac{\Phi_0^2}{r^2}-\ii q\kappa l(l+1)\int^\infty_0\dd r \frac{\Phi_0 \mathcal A}{r}. 
\end{equation}
Then the imaginary part of $\omega_1$ can be calculated as 
\begin{eqnarray}
{\rm Im}\omega_1&=&-q\kappa \int^\infty_0 \dd \xi \frac{\Phi_0(\xi)}{\xi} {\rm Re} \mathcal A(\xi)
=q^2\kappa^2\omega_0\left(\int^\infty_0 \dd \xi \frac{\Phi_0(\xi)}{\xi}j_1(\xi)\right)^2\cr
&=&\frac{q^2\kappa^2}{96a_0^3}\frac{\left(2a_0\omega_0-{\rm ArcTan}(2a_0\omega_0)\right)^2}{a_0^4\omega_0^4}
\sim \frac{1}{24}\frac{q^2\kappa^2}{a_0^2}\frac{1}{a_0^3\omega_0^3}\omega_0\sim \frac{1}{24}\mu^2 q^2\kappa^2(GM\mu)^5\mu,   
\end{eqnarray}
where we estimated the value of $\omega_0$ as $\omega_0\sim\mu$ in the last expression. 
Therefore the imaginary part is positive, and the amplitude of $\phi$ decays. 
%

\section{Analysis with the uniform magnetic field}
\label{sec:wald}
\subsection{background and perturbation equations}
Let us consider the magnetic field in the Schwarzschild geometry given by~\cite{Wald:1974np} 
\begin{equation}
F=B_0r\sin^2\theta \dd r\wedge \dd \varphi +B_0 r^2\sin\theta \cos\theta \dd\theta\wedge \dd \varphi, 
\end{equation}
where $F$ is the two-form field strength. 
By using the radial coordinate $\rho:=r\sin\theta$ of the cylindrical coordinates, 
we obtain 
\begin{equation}
F=B_0 r \sin\theta \dd\rho\wedge \dd\varphi=B_0 e_\rho\wedge e_\varphi. 
\end{equation}
Therefore the field strength describes the uniform magnetic field along the $z$-direction. 
The vector potential is given by 
\begin{equation}
A_\mu^{\rm bg}=\frac{1}{2}B_0 r^2\sin^2\theta (\dd\varphi)_\mu. 
\end{equation}
This field configuration satisfies the field equations with $\phi=0$. 
Let us consider the following perturbation: 
\begin{eqnarray}
\phi&=&\delta \phi,\\
A^{\rm tot}_\mu&=&\frac{1}{2}B_0 r^2\sin^2\theta (\dd\varphi)_\mu+\delta A_\mu.  
\end{eqnarray}
Hereafter we will take the gauge $A^c_{lm}=0$~\cite{Zerilli:1974ai} as in the previous section. 

Since the background magnetic field violates the spherical symmetry, 
multiple modes are coupled with each other. 
Therefore the scalar field equation is given by Eq.~\eqref{eq:fullphieom} with the summation symbol, that is, 
\begin{equation}
\sum_{lm}\left\{\left[\frac{f}{r^2}\del_r\left(r^2f\del_r \Phi_{lm}\right)+\omega^2\Phi_{lm}
-f\mu^2\Phi_{lm}-f\frac{l(l+1)}{r^2}\Phi_{lm}\right]Y_{lm}\right\}=\frac{1}{4}\kappa f F_{\mu\nu}\tilde F^{\mu\nu}, 
\label{eq:fullphieom_uni}
\end{equation}
where 
\begin{eqnarray}
F_{\mu\nu}\tilde F^{\mu\nu}&=&\frac{4B_0}{r}\left[r\cos\theta\left(\del_r\delta A_t-\del_t \delta A_r\right)+\sin\theta\left(-\del_\theta \delta A_t+\del_t \delta A_\theta\right)\right]\cr 
&=&\frac{4\ii B_0}{r}\sum_{lm}\Biggl[-r\cos\theta\left(\del_r A^a_{lm}+\omega A^b_{lm}\right)Y_{lm}
\cr&&\hspace{3cm}+
\sin\theta
A^a_{lm}
\del_\theta Y_{lm}
+ 
\frac{\omega}{\sqrt{l(l+1)}}A^d_{lm}\del_\varphi Y_{lm}\Biggr]. 
\end{eqnarray}
The gauge field equations are given by 
\begin{eqnarray}
&&\sum_{l,m}\left\{ \left[
f\del_r(r^2\del_r A^a_{lm})+\omega f \del_r(r^2 A^b_{lm})-l(l+1)A^a_{lm}\right]Y_{lm}\right\}
\cr
&&\hspace{5cm}
=\ii B_0\kappa rf\sum_{l,m}\left(r\del_r\Phi_{lm}\cos\theta Y_{lm}-\Phi_{lm}\sin\theta\del_\theta Y_{lm}\right),
\label{eq:t}
\\
&&\sum_{l,m}\left\{ \left[
\omega r^2\del_r A^a_{lm}+\omega^2r^2A^b_{lm}-l(l+1)fA^b_{lm}\right]Y_{lm}\right\}
\cr
&&\hspace{5cm}
=\ii B_0\kappa \omega r^2 \cos\theta \sum_{l,m}\Phi_{lm}Y_{lm},~~
\label{eq:r}
\\
&&\sum_{l,m}\Bigl\{ 
r^2 \del_\theta Y_{lm}\left[
\omega A^a_{lm}-f\del_r(fA^b_{lm})\right]
\cr
&&\hspace{1.5cm}
+
\frac{\del_\varphi Y_{lm}}{\sqrt{l(l+1)}\sin \theta}
\left[r^2f\del_r(f\del_r A^d_{lm})+\omega^2 r^2 A^d_{lm}-l(l+1)fA^d_{lm}\right] \Bigr\}
\cr
&&\hspace{5cm}
=-\ii B_0\kappa \omega r^3 f \sin\theta\sum_{lm}\Phi_{lm}Y_{lm},
\label{eq:th}
\\
&&\sum_{l,m}\Biggl\{ 
r^2  \del_\varphi Y_{lm}\left[
\omega A^a_{lm}-f\del_r(fA^b_{lm})\right]
\cr
&&\hspace{1.5cm}
-
\frac{\sin\theta\del_\theta Y_{lm}}{\sqrt{l(l+1)}}
\left[r^2f\del_r(f\del_r A^d_{lm})+\omega^2 r^2 A^d_{lm}-l(l+1)fA^d_{lm}\right] \Biggr\} 
=0. 
\label{eq:ph}
\end{eqnarray}

Combining Eq.~\eqref{eq:th} and Eq.~\eqref{eq:ph}, we obtain the following equation for $A^d_{lm}$:
\begin{equation}
\sum_{l,m}\left\{\sqrt{l(l+1)}\left[r^2f\del_r(f\del_r A^d_{lm})+\omega^2r^2 A^d_{lm}-l(l+1)fA^d_{lm}\right]Y_{lm}\right\}=
\ii B_0\kappa \omega r^3 f \sum_{lm}\Phi_{lm} Y_{lm}. 
\label{eq:uniodd}
\end{equation}
We can obtain the other independent equation from Eq.~\eqref{eq:th} and Eq.~\eqref{eq:ph} as follows:
\begin{equation}
\sum_{l,m}\left\{l(l+1)\left[
\omega A^a_{lm}-f\del_r (fA^b_{lm})\right]Y_{lm}\right\}
=\ii B_0 \kappa \omega r \sum_{lm}\Phi_{lm}\left(2\cos\theta Y_{lm}+\sin\theta \del_\theta Y_{lm}\right). 
\label{eq:tr}
\end{equation}
Therefore $A^a_{lm}$ and $A^b_{lm}$ are given by the coupled equations \eqref{eq:r} and \eqref{eq:tr}.

\subsection{Approximate Analysis under the condition \eqref{eq:approx}}

Let us impose the condition \eqref{eq:approx} and 
\begin{equation}
\kappa B_0 \ll 1/a_0 
\label{eq:smallB}
\end{equation}
so that we can consider the right-hand side of Eq.~\eqref{eq:fullphieom_uni} is perturbatively small. 
Then we can assume that 
the dominant part of the axion field is given by 
the $n=2$, $l=1$ and $m=\pm 1$ mode of the solution for Eq.~\eqref{eq:Hydro} 
as in the previous section, 
where we have assumed that the uniform magnetic field is parallel~($m=1$) or anti-parallel~($m=-1$) to 
the dominant mode. 
In the context of the superradiant instability, the magnetic field is parallel~(anti-parallel) to the 
angular momentum of the black hole for $m=1$~($m=-1$). 
The dominant part of the axion field can be explicitly written as 
\begin{equation}
\delta \phi_0=\Phi_0(r)\ee^{\ii\omega t}Y_{1~\pm1}(\theta, \phi)=\mp\sqrt{\frac{3}{8\pi}}\sin\theta \ee^{\ii\omega t\pm\ii \varphi}\Phi_0(r), 
\end{equation}
where $\Phi_0$ is given in Eq.~\eqref{eq:phi21}. 

From Eq.~\eqref{eq:uniodd}, we find that 
$A^d_{1~\pm 1}$ is induced by $\delta\phi_0$ through the following equation:
\begin{equation}
r^2f\del_r(f\del_r A^d_{1~\pm 1})+\omega^2r^2 A^d_{1~\pm 1}-2fA^d_{1~\pm 1}=
\mp\frac{1}{\sqrt{2}}B_0\kappa \omega r^3 f \Phi_0 . 
\label{eq:Ad}
\end{equation}
From Eqs.~\eqref{eq:r} and \eqref{eq:tr}, 
having the expression 
\begin{equation}
Y_{2~\pm1}=\mp\sqrt{\frac{15}{8\pi}}\sin\theta\cos\theta \ee^{\pm\ii \varphi}, 
\end{equation}
we find that only $A^{a,b}_{2~\pm1}$ modes 
can be induced by $\delta\phi_0$. 
Then, eliminating $A^a_{2~\pm1}$ from Eqs.~\eqref{eq:tr} and \eqref{eq:r} with $l=2$ and $m=\pm1$, 
we obtain
\begin{equation}
-6fA^b_{2~\pm1}+r^2\del_r(f\del_r (f A_{2~\pm1}^b))+r^2\omega^2 A^b_{2~\pm1}
=\frac{\ii}{2\sqrt{5}}B_0\kappa \omega r^2\left[2\Phi_0-\del_r (rf\Phi_0)\right], 
\label{eq:Ab}
\end{equation} 
and $A^a_{2~\pm 1}$ is given by 
\begin{equation}
A^a_{2~\pm1}=\frac{f}{\omega}\del_r(fA^b_{2~\pm1})+\frac{\ii}{2\sqrt{5}}B_0\kappa f r \Phi_0. 
\end{equation}

Introducing $\mathcal B_\pm:=A^b_{2~\pm1}/(\omega_0 r)$ and $\mathcal D_\pm:=A^d_{1~\pm1}/r$,  and assuming the condition \eqref{eq:approx}, 
we obtain 
\begin{eqnarray}
 &&\frac{\dd}{\dd x}\left(x^2\frac{\dd}{\dd x}\mathcal B_\pm\right)+x^2\mathcal B_\pm-6\mathcal B_\pm=\ii\frac{B_0\kappa}{2\sqrt{5}\omega_0}x\left(\Phi_0-x\frac{\dd}{\dd x}\Phi_0\right),
\label{eq:B}\\ 
&&\frac{\dd}{\dd x}\left(x^2\frac{\dd}{\dd x}\mathcal D_\pm\right)+x^2\mathcal D_\pm-2\mathcal D_\pm=\mp\frac{B_0\kappa}{\sqrt{2}\omega_0}x^2\Phi_0 
\label{eq:D}
\end{eqnarray}
from Eqs.~\eqref{eq:Ad} and \eqref{eq:Ab}. 
Eqs.~\eqref{eq:B} and \eqref{eq:D} can be solved by using the Green's function \eqref{eq:green}. 
As in the case of Eq.~\eqref{eq:reA},  
the functions ${\rm Re} \mathcal B_\pm$ and ${\rm Im} \mathcal D_\pm$ can be expressed as 
\begin{eqnarray}
{\rm Re} \mathcal B_\pm(x)&=& \frac{B_0\kappa}{2\sqrt{5}\omega_0}j_2(x)\int^\infty_0 \dd\xi \xi j_2(\xi)\left(\xi \del_\xi \Phi_0(\xi)-\Phi_0(\xi)\right), \\
{\rm Im} \mathcal D_\pm(x)&=&\mp\frac{B_0\kappa}{\sqrt{2}\omega_0}j_1(x)\int^\infty_0 \dd\xi \xi^2 j_1(\xi) \Phi_0(\xi). 
\end{eqnarray}

For $F_{\mu\nu}\tilde F^{\mu\nu}$, we obtain 
\begin{eqnarray}
F_{\mu\nu}\tilde F^{\mu\nu}
&=&\frac{4\ii B_0}{r}\Biggl[-r\left(\del_r A^a_{2~\pm1}+\omega A^b_{2~\pm1}\right)\cos\theta Y_{2~\pm1}
+
A^a_{2~\pm1}\sin\theta \del_\theta Y_{2~\pm1}
\pm
\frac{\ii\omega}{\sqrt{2}}A^d_{1~\pm1}Y_{1~\pm1}\Biggr]\cr
&=&4\ii B_0\Biggl\{\left[-\frac{1}{\sqrt{5}}\left(\del_r A^a_{2~\pm1}+\omega A^b_{2~\pm1}
+\frac{3}{r}A^a_{2~\pm1}\right)\pm\frac{\ii}{\sqrt{2}r}\omega A^d_{1~\pm1}\right]Y_{1~\pm1}
\cr&&\hspace{1cm}
+\sqrt{\frac{8}{35}}\left(-\del_r A^a_{2~\pm1}-\omega A^b_{2~\pm1}+\frac{2}{r}A^a_{2~\pm1}\right)Y_{3~\pm1}\Biggr\}. 
\end{eqnarray}
Taking the same approximation \eqref{eq:approx}, from the axion field equation \eqref{eq:fullphieom_uni} with $l=1$, 
we obtain the equation for $\Phi_1$ and $\omega_1$ as follows:
\begin{eqnarray}
&&\frac{\dd^2 \Phi_1}{\dd r^2}+\frac{2}{r}\frac{\dd \Phi_1}{\dd r}+\left(\omega_0^2-\mu^2\right)\Phi_1
+2\omega_0\omega_1\Phi_0
+\frac{2}{a_0r}\Phi_1-\frac{2}{r^2}\Phi_1\cr
&&\hspace{2cm}
=\ii\kappa B_0 \left[-\frac{1}{\sqrt{5}}\left(\del_r A^a_{2~\pm1}+\omega_0 A^b_{2~\pm1}+\frac{3}{r}A^a_{2~\pm1}\right)\pm\frac{\ii}{\sqrt{2}r}\omega_0 A^d_{1~\pm1}\right].
\end{eqnarray}
Multiplying $r^2\Phi_0$ and integrating by parts, we obtain
\begin{equation}
2\omega_0\omega_1\int^\infty_0 \dd r r^2 \Phi_0^2=\ii \kappa B_0 
\int^\infty_0 \dd r r^2 \Phi_0\left[-\frac{1}{\sqrt{5}}\left(\del_r A^a_{2~\pm1}+\omega_0 A^b_{2~\pm1}+\frac{3}{r}A^a_{2~\pm1}\right)\pm\frac{\ii}{\sqrt{2}r}\omega_0 A^d_{1~\pm1}\right]. 
\end{equation}
Focusing on the imaginary part of $\omega_1$ and integrating by parts, we obtain
\begin{equation}
{\rm Im}\omega_1=\frac{\kappa B_0}{2\omega_0^2} 
\int^\infty_0 \dd x \left[\frac{3}{\sqrt{5}} x\left(x\del_x \Phi_0-\Phi_0\right){\rm Re}\mathcal B_\pm \mp\frac{1}{\sqrt{2}}x^2\Phi_0 {\rm Im}\mathcal D_\pm\right]. 
\end{equation}
Then ${\rm Im}\omega_1$ is given by 
\begin{eqnarray}
{\rm Im}\omega_1&=&\frac{\kappa^2B_0^2}{4\omega_0^3}\left[\frac{3}{5}\left( \int^\infty_0 \dd\xi \xi j_2(\xi)\left(\xi \del_\xi \Phi_0(\xi)-\Phi_0(\xi)\right)\right)^2+\left(\int^\infty_0 \dd\xi \xi^2 j_1(\xi) \Phi_0(\xi)\right)^2
\right] \cr
&=&\frac{2^{14}}{15} a_0^2 \kappa^2 B_0^2 \frac{a_0^3 \omega_0^3}{(1+4 a_0^2\omega_0^2)^6}\omega_0\sim \frac{4}{15} a_0^2 \kappa^2 B_0^2 \frac{1}{a_0^9\omega_0^9}\omega_0\sim\frac{4}{15}\frac{\kappa^2 B_0^2}{\mu^2}(GM\mu)^7 \mu. 
\end{eqnarray}

\section{Comparison with the growth rate of the superradiant instability}
\label{sec:comp}

Let us compare the decay rates obtained in Sec.~\ref{sec:mono} and Sec.~\ref{sec:wald} 
with the growth rate $\omega_{\rm sr}$ of the superradiant instability given in Refs.~\cite{Detweiler:1980uk,Pani:2012bp} 
as follows:
\begin{equation}
\omega_{\rm sr}\sim \frac{1}{48}(GM\mu)^8\mu\sim
2\times 10^{-17}{\rm s}^{-1}\left(\frac{\mu}{10^{-18}{\rm eV}}\right)^9 \left(\frac{M}{4\times 10^6 M_\odot}\right)^8. 
\end{equation}
The ratio between $\omega_{\rm sr}$ and ${\rm Im}\omega_1$ is given by 
\begin{eqnarray}
\left(\frac{{\rm Im}\omega_1}{\omega_{\rm sr}}\right)_{\rm mono}&\sim &
2\left(\frac{\kappa^2q^2}{a_0^2}\right)\times (GM\mu)^{-5}, 
\label{eq:ratiomono}
\\
\left(\frac{{\rm Im}\omega_1}{\omega_{\rm sr}}\right)_{\rm uni}&\sim& 
\frac{64}{5}(a_0^2\kappa^2B_0^2)\times (GM\mu)
\label{eq:ratio}
\end{eqnarray}
for the monopole magnetic field and the uniform magnetic field, respectively. 
We note that we have assumed $\kappa q/a_0 \ll 1$, $a_0\kappa B_0\ll 1$ and $GM\mu\ll 1$ in our 
analytic evaluation. 
Within the relevant parameter region of our approximation, 
the ratio can be much larger than unity if the axion mass is sufficiently smaller than $1/(GM)$ 
for the monopole magnetic field. 
On the other hand, the ratio cannot be 
much larger than unity for 
the uniform magnetic field within the relevant parameter region of our approximation. 
Nevertheless, it would be reasonable to expect that a similar decaying process exists even if 
$a_0 \kappa B_0\sim 1$ and/or $GM\mu\sim 1$. 

Applying the formulae \eqref{eq:ratiomono} and \eqref{eq:ratio} to the black hole at the center of our galaxy, 
we find
\begin{eqnarray}
&&\left(\frac{{\rm Im}\omega_1}{\omega_{\rm sr}}\right)_{\rm mono}\sim 
2\mu^2\kappa^2q^2(GM\mu)^{-3}\cr
&&\hspace{1cm}\sim 
3\times 10^8 \left(\frac{\kappa}{10^{-12}{\rm GeV}^{-1}}\right)^2
\left(\frac{q\mu^4 (GM)^2 }{10^3{\rm G}}\right)^2
\left(\frac{\mu}{10^{-18}{\rm eV}}\right)^{-9}
\left(\frac{M}{4\times 10^6 M_\odot}\right)^{-7}, 
\\
&&\left(\frac{{\rm Im}\omega_1}{\omega_{\rm sr}}\right)_{\rm uni}\sim 
\frac{64}{5}\frac{\kappa^2B_0^2}{\mu^2}\frac{1}{GM\mu}\cr
&&\hspace{1cm}\sim 
2 \left(\frac{\kappa}{10^{-12}{\rm GeV}^{-1}}\right)^2
\left(\frac{B_0}{10^3{\rm G}}\right)^2
\left(\frac{\mu}{10^{-18}{\rm eV}}\right)^{-3}
\left(\frac{M}{4\times 10^6 M_\odot}\right)^{-1}, 
\end{eqnarray}
where we referred to Ref.~\cite{Berg:2016ese} for the value of $\kappa$.

\section{summary and discussion}

We derived the decay rate of the axion cloud around a black hole 
considering the axion-photon coupling with background magnetic fields. 
Although the axion field has unstable bound states around a rotating black hole 
without the electro-magnetic fields, 
the axion-photon coupling together with a background magnetic field 
induces the axion-photon conversion, and the photon can dissipate to infinity. 
We obtained analytic expression of the decay rate for the monopole and uniform magnetic field configurations 
under the approximation for which 
the axion configuration is dominated by the solution of the equation equivalent to 
that of the Hydrogen atom. 
For the monopole background, we found 
the decay rate ${\rm Im}~\omega_1\sim \mu^2 q^2\kappa^2(GM\mu)^5\mu/24$, 
where $\kappa$, $q$, $\mu$ and $M$ are the axion-photon coupling constant, magnetic charge, 
axion mass and black hole mass, respectively. 
For the uniform background magnetic field, we obtain 
the decay rate as $\sim \kappa^2 B_0^2 \mu^6 (GM)^7$, 
where $B_0$ is the magnetic field strength. 
Since the two expressions for the monopole background and the uniform magnetic field are 
different from each other, the decay rate may significantly depend on 
the configuration of the background magnetic field. 

When we discuss an axion cloud around a black hole, 
the decay rate should be sufficiently small compared to the 
growth rate of the superradiant instability and the inverse of the relevant time scale. 
Let us consider the central black hole of our galaxy, Sgr A* as a specific example. 
The ambient environments of Sgr A* will be probed with S-stars~\cite{Amorim:2019hwp} and possible pulsars 
orbiting around Sgr A* which are expected to be discovered by SKA~\cite{Liu:2011ae}. 
Applying our formulae, we found that the magnitude of the decay rate due to the axion-photon conversion 
can be comparable to that of the growth rate of the superradiant instability 
for the central black hole in our galaxy. 
Therefore Sgr A* is one of notable systems which could be relevant to the phenomenon we described in this paper. 

It should be noted that, in the analysis, we have imposed several conditions to perform the analytic evaluation. 
Moreover, the frequency of the photon, which is comparable to the axion mass scale in the present case, 
would be much lower than the plasma frequency in the black hole environment, and the photon emission would be prohibited. 
Nevertheless, one would expect that an alternative wave mode such as Alfv\'en waves may play 
a similar role to the photon and a similar decay process may 
generally exist. 
Obviously, in order to clarify the decay process, we need to investigate more realistic situations 
possibly realized in a black hole system. 
These are our future issues.

\section*{Acknowledgements}
We thank Jiro Soda for helpful comments and his excellent lecture on axion physics, 
which was publicly open and remotely provided as an intensive lecture course at Rikkyo University.  
We also thank Takahiro Tanaka for helpful comments.  
This work was supported by JSPS KAKENHI Grant
Numbers JP19H01895~(C.Y.), JP20H05850~(C.Y.), JP20H05853~(C.Y.), 20H05852~(A.N.), JP19H01891~(A.N. and D.Y.), 
15H05896~(K.T.), 16H05999~(K.T.), 17H01110~(K.T.), 20H00180~(K.T.) and 17K14304~(D.Y.), 
Bilateral Joint Research Projects of JSPS~(K.T.), and the ISM Cooperative Research Program [2020-ISMCRP-2017]~(K.T.).



\end{document}